%% file: main.tex
\documentclass[conference]{IEEEtran}
\usepackage{cite}
\usepackage{amsmath,amssymb,amsfonts}
\usepackage[ruled,linesnumbered,vlined]{algorithm2e}
\usepackage{graphicx}
\usepackage{textcomp}
\usepackage{xcolor}
\usepackage{booktabs}
\usepackage{kotex}
\usepackage[normalem]{ulem}
\usepackage{subfigure}
\usepackage{adjustbox}

\def\BibTeX{{\rm B\kern-.05em{\sc i\kern-.025em b}\kern-.08em
    T\kern-.1667em\lower.7ex\hbox{E}\kern-.125emX}}

\begin{document}

\title{Effectively Sampling Higher Order Mutants Using Causal Effect}

\input{Author}
\maketitle
\input{Abstract}
\input{Introduction}
\input{Motivating_Example}
\input{cpda}
\input{HUE}
\input{Evaluation}
\input{Result}
\input{related_work}
\input{Conclusion}

\bibliographystyle{IEEEtran}
\bibliography{ref}

\end{document}

%% file: Author.tex

\author{\IEEEauthorblockN{Saeyoon Oh}
\IEEEauthorblockA{\textit{School of Computing} \\
\textit{KAIST}\\
Daejeon, Republic of Korea \\
saeyoon17@kaist.ac.kr}
\and
\IEEEauthorblockN{Seongmin Lee}
\IEEEauthorblockA{\textit{School of Computing} \\
\textit{KAIST}\\
Daejeon, Republic of Korea \\
bohrok@kaist.ac.kr}
\and
\IEEEauthorblockN{Shin Yoo}
\IEEEauthorblockA{\textit{School of Computing} \\
\textit{KAIST}\\
Daejeon, Republic of Korea \\
shin.yoo@kaist.ac.kr}
}

%% file: Abstract.tex

\begin{abstract}
Higher Order Mutation (HOM) has been proposed to avoid equivalent mutants and 
improve the scalability of mutation testing, but generating useful HOMs remain 
an expensive search problem on its own. We propose a new approach to generate 
Strongly Subsuming Higher Order Mutants (SSHOM) using a recently introduced 
Causal Program Dependence Analysis (CPDA). CPDA itself is based on program 
mutation, and provides quantitative estimation of how often a change of the 
value of a program element will cause a value change of another program 
element. Our SSHOM generation approach chooses pairs of program elements using 
heuristics based on CPDA analysis, performs First Order Mutation to the chosen 
pairs, and generates an HOM by combining two FOMs. 
\end{abstract}

%% file: Introduction.tex

\section {Introduction}
\label{sec:introduction}

Mutation testing aims to inject artificial faults (i.e., mutate the program 
under test) to evaluate the adequacy of the existing test 
suite~\cite{Jia:2011nx}. Since the evaluation of test adequacy is based on 
concrete detection of actual faults (i.e., syntactic changes made to the 
program under test), mutation testing has great potential to improve the 
effectiveness of software testing. However, a couple of problems limit the 
practical applicability of mutation testing: the high cost of mutation 
analysis, in which we generate, build, and test a large number of mutants, and 
the existence of equivalent mutants, whose detection is theoretically 
undecidable.

Higher Order Mutation, in which multiple mutations are combined to create 
a Higher Order Mutant (HOM), has been proposed as a solution to the equivalent 
mutant problem.  Through an empirical study, Offutt showed that approximately 
10\% of First Order Mutants (FOMs) are equivalent, while only about 1\% of the 
second order mutants are equivalent~\cite{Offutt:1992}.

However, the number of possible HOMs is far greater than that of FOMs, due to 
the combinatorial nature of HOMs. This exacerbates the problem of efficiency 
and scalability. While it is possible to devise a combinatorial approach for 
generation and sampling of HOMs~\cite{Madeyski2014}, a purely combinatorial 
approach may fall short because HOMs are by definition easier to kill (i.e., 
they deviate more from the original program due to multiple mutations). A HOM 
that is trivial to kill may not be beneficial. 

The concept of Strongly Subsuming Higher Order Mutant (SSHOM) was introduced 
by Jia and Harman~\cite{Jia2009id}. SSHOMs are higher order mutants that are 
harder to kill than its constituent first order mutants. If it is possible to  
search for SSHOMs efficiently, it may reduce the number of HOMs we need to 
consider. A search based approach has been studied for the generation of SSHOMs~\cite{Harman14ase}: the fitness function guides the search towards 
satisfying the conditions of SSHOMs.

In this paper, we propose constructive heuristics for generating SSHOMs based 
on a novel type of dependence analysis called Causal Program Dependence 
Analysis (CPDA)~\cite{Lee:2021aa}. Based on CPDA, we propose two heuristics that are designed to 
sample SSHOMs not only more efficiently, but also in a way that produces 
diverse mutants. 

Unlike traditional static dependence analysis that determines whether a 
program element depends on another element in a Boolean fashion, CPDA assigns 
quantitative likelihood values to each dependence: higher likelihood means 
that the change of value in one element is highly likely to cause the change 
in the value of another element. Using this, we can prioritise pairs of 
program elements in the order of the likelihood of FOMs generated at each 
location masking each other when combined as a HOM, in turn resulting in a 
SSHOM. We propose heuristics that will choose different pairs of program 
elements based on the results of CPDA. A case study with two small, easy to 
analyse programs suggests that CPDA based generation of SSHOMs can be effective.

The remainder of this paper are as follows. In Section~\ref{sec:motivating_example}, we show how dependency between program elements can be used to generate strongly subsuming higher order mutants, using a motivating example. Section~\ref{sec:cpda} introduces Causal 
Program Dependence Analysis, an approach to calculate dependence. Section~\ref{sec:heuristics} 
introduces four heuristics for sampling higher order mutants. Section~\ref{sec:experimental_setup} describes the setup of our evaluation, and Section~\ref{sec:results} presents the results. Section~\ref{sec:related_work} presents related work, and Section~\ref{sec:conclusion} concludes.

%% file: Motivating_Example.tex

\section{A Motivating Example}
\label{sec:motivating_example}

Consider a HOM $h$ that is composed of two FOMs, $f_1$ and $f_2$. In order for 
$h$ to be a SSHOM, there must exist at least one test, $t$, such that $t$ 
fails due to $f_1$ as a FOM, which is masked by $f_2$, so that $t$ does not 
fail under $h$. For the masking to take place, it is 
reasonable to assume that the locations of $f_1$ and $f_2$ are connected by 
program dependence: otherwise, a mutation in one of the location is not likely 
to mask the mutation in another.

\begin{algorithm}[ht]
    \caption{Example Program\label{alg:example}}
    \DontPrintSemicolon
    \SetKwProg{Main}{Function}{:}
    
    \Main{Main (b: int)}{
        $a = 1$
     
        $a = a + 1$
        
        \If{$b~\%~2 = 0$}
            {$a = a \times 2$}
        
        $c = 100$
        
        $return$ $a$
    }
\end{algorithm}

Algorithm~\ref{alg:example} contains a concrete motivating example. We are 
going to use Expression Replacement mutation operator to generate HOMs. Let 
Line 2 be one of the constituent FOMs: for example, we mutate it to $a = 2$. 
Now, let us consider the program element in Line 3. Whenever $a$ in Line 2 
gets mutated, the effect is delivered to program element in Line 3. In other 
words, program element in Line 3 is \emph{frequently} dependent on program 
element in Line 2. In contrast, the variable $a$ in Line 5 is affected by
$a$ in Line 2 only when $b$ is even. The element in Line 5 is, therefore, 
\emph{less frequently} dependent on element in Line 2 when compared to 
$a$ in Line 3. Finally, program element in Line 6 is not dependent on $a$ 
in Line 2.

Now consider mutating either Line 3, 5, or 6, to mask the effect of the mutation 
in Line 2 towards Line 7. 
Mutating program element in Line 3 has highest chance of masking the first 
mutation, since the effect of the mutation in Line 2 is always delivered to 
Line 3. Mutating Line 5 \emph{can} mask the fault, but with a smaller 
probability, since the effect of the mutation in Line 2 is only delivered when 
$b$ is even. Finally, no mutation in Line 6 can mask the mutation in Line 2, 
since it cannot change the value of $a$.

Through the motivating example, we can observe that the more \emph{frequently} 
dependence relation occurs between two program elements, the higher the 
chance of fault masking will be. The fact that we need program dependence 
relationship between two program elements for fault masking to happen may be a 
trivial observation. However, what Causal Program Dependence Analysis allows 
us to reason about is the relative likelihood of a dependence relationship 
actually affecting the value of a specific program element. Unlike traditional 
dependence analysis whose outcome is binary (either dependent or not), CPDA 
allows us to reason about the degree of dependence quantitatively. We will 
present a brief introduction to CPDA in Section~\ref{sec:cpda}.

%% file: cpda.tex

\section{Causal Program Dependence Analysis}
\label{sec:cpda}

\emph{Causal Program Dependence Analysis} (CPDA), a recently introduced dynamic program dependence analysis technique, can measure the degree of dependence between two program elements~\cite{Lee:2021aa}. Applying the causal inference~\cite{Pearl:2009aa} on the program execution trajectory, CPDA estimates how often a change of the value of a program element causes a change of the value of another program element.

CPDA calculates the dependence by a given test suite. CPDA first generates data on which program element's values are associated. It gets association data by running tests on programs that have modified part of the code in various ways, observing which program elements have simultaneously changed compared to the values they had in the original program. Given the association data, CPDA discovers the \emph{causal structure} of the program. The causal structure is a directed graph that represents the direct dependence between program elements; for each child node in the causal structure, the set of parent nodes (immediate predecessor nodes) comprises a minimal Markov blanket of the behavior of the child node~\cite{Pearl:2009aa}. Using the association data and discovered causal structure, CPDA estimates two metrics representing the degree of dependence. A \emph{causal effect} is an aggregate of the effect of each program element's change causing a change in other program elements. A \emph{direct effect} is the effect of one program element on another, excluding all the indirect effects through other program elements.

Given association data $O$, the causal effect from a program element $S_i$ to the other program element $S_j$, denoted as $\mathit{CE}_O(S_1, S_j)$, is calculated as follows:
\begin{align*}
\label{def:ce}
    \mathit{CE}_O(S_i, S_j) &= P_O(S_j = 1 \mid do(S_i = 1)) \\ &\times (1 - P_O(S_j = 1 \mid do(S_i = 0))).
\end{align*}
. In the equation, $S_j = 1$ implies the value of the program element $S_j$ is changed compared to the original program, or otherwise, $S_j = 0$. $P_O(x \mid do(y))$ calculates the probability of an event $x$ \emph{caused} by the event $y$~\cite{Pearl:2009aa}. The causal effect aims to measure the difference of the effect that $S_j$ gets when $S_i$ moves its state from \emph{unchanged} (0) to \emph{changed} (1). Instead of subtracting the probability, we multiply the probability when $S_i = 1$ and the complementary probability when $S_i = 0$, keeps the causal effect having a positive value.

From the example in Sec.~\ref{sec:motivating_example}, SSHOMs can be created more easily from pairs of program locations that affect more frequently. Thus, we claim that we could efficiently search the higher-order mutant space to sample SSHOMs by utilizing the CPDA. Our main hypothesis is that the causal effect could prioritize the second-order mutation position for SSHOMs. Our empirical evaluation investigates whether there is a positive correlation between the high causal effect and the strongly subsuming rate. Assuming the hypothesis is plausible, setting up an actual guideline for searching on the higher-order mutant space is needed. In the next section, we introduce several heuristics that can efficiently sample the higher-order mutant using the causal effect.

%% file: HUE.tex

\section{Heuristics for SSHOM Generation}
\label{sec:heuristics}

In this section, we introduce heuristics to effectively sample second order mutants (SOM).

\subsection{Heuristics for selecting SOM}

In order to search for efficient methods of sampling second order mutants we 
came up with four heuristics: \emph{Random}, \emph{Prop}, 
\emph{Dsort}, and \emph{MWM}. The purpose of each heuristic is to 1) sample as 
much SSHOM as possible with fixed number of mutants, 2) sample diverse 
mutants. Therefore, each heuristic tries to find the best pair of program 
elements to generate the second order mutants along with the number of mutants 
to generate per pair. 
While we investigate through second order mutants, it can easily be extended
to select three for more mutation locations utilizing the causal effect (e.g., 
choosing $n$ locations whose sum of the pairwise causal effect is high.)

All heuristics other than Random approach make use of the dependency 
calculated in terms of causal effect. Therefore, modeling CPDA and 
calculating the causal effect comes firsthand for the three last heuristics. 

The four heuristics are as 
follows.

\subsubsection{Random}

Among all possible pairs of program elements, we choose a random pair. We repeat 
this process independently for number of total mutants, while allowing 
duplicates. If a certain pair is selected $k$ times, we create $k$ mutants by 
mutating the pair of program elements.

\subsubsection{Prop (Proportional)}

We first weight each pair of program elements with their causal effect 
values. Subsequently, we choose a pair with a probability proportional to its 
weight. Let $\mathit{PP}$ be the set of all pairs of program elements. Let $CE(p_i)$ be the causal effect value of $p_i \in \mathit{PP}$. The probability of $p_i$ being selected, $P(p_i)$ is:

\begin{equation}\label{eq:proportional}
    P(p_i) = \frac{CE(p_i)}{\sum_{p_j \in \mathit{PP}} {CE(p_j)}}
\end{equation}

Note that, according to Equation~\ref{eq:proportional}, a pair with causal 
effect value of 0 cannot be selected. We repeat this process of choosing a 
pair for number of total mutants. Similar to Random, if a pair of program 
elements is selected $k$ times, we build $k$ mutants by mutating the pair.

\subsubsection{Dsort (Descending Sort)}

With Dsort, we sort all pairs of program elements according to their causal 
effect values, and choose the top $n$ pairs. Subsequently, we distribute the 
number of mutants to generate, $k$, equally to the chosen $n$ pairs. In our 
evaluation, we set $n$ for Dsort as the number of pairs selected by the MWM heuristic. 

\subsubsection{MWM (Maximum Weight Matching)}
While Dsort picks program pairs by considering solely causal effect, 
MWM considers the diversity of the mutant set. Since the purpose of each 
mutant is to mimic real faults of developers a good mutant set should contain 
diverse mutants rather than mutants with similar faults. 

To sample diverse set of mutants, we utilize \emph{maximum weight matching} from graph theory. 
A set $M$ of independent edges in a graph $G = (V, E)$ is called a 
matching~\cite{Diestel:2016aa}. A maximum weight matching $M$ of graph 
$G = (V, E)$ where every edge $e \in E$ have weight $w_e$ is set $M$ of 
independent edges maximizing the sum of weights of edges in $M$.


To perform maximum weight matching with respect to causal effects in the program,
we modify the causal structure $G$. 
Whenever there is a path from $v_1$ to $v_2$, we add a directed edge from $v_1$ to $v_2$.
Subsequently, we weight all edges: the weight 
of $e$ from $v_1$ to $v_2$ is the causal effect $v_2$ gets from $v_1$.

After modifying the causal structure $G$ we compute the maximum weight 
matching of $G$. The aim is not only to select pairs with high dependency, but 
also to select a diverse set of program pairs across the entire program. 
Similar to Dsort, total number of mutants to be made is then distributed 
equally to all program pairs, making same number of second order mutants per 
pair.

%% file: Evaluation.tex

\section{Experimental Setup}
\label{sec:experimental_setup}

\subsection{Research Questions}

\subsubsection*{RQ1. Causal Effect and SSHOM} Does high Causal Effect lead to 
strongly subsuming second order mutant? To answer this question, we first 
calculate causal effects between program elements of studied programs. We then 
distribute the pairs with non-zero causal effect values into ten equal size 
buckets; we also group all pairs with causal effect value of 0 in a separate 
bucket. Subsequently, we randomly select five program element pairs from 
each of the 11 buckets, and generate 100 second order HOMs from each pair. We 
then calculated the number of SSHOM made from each bucket. We repeat this ten 
times to remove the sampling bias.

\subsubsection*{RQ2. SSHOM Heuristics} How do different heuristics compare to 
each other in terms of the number of SSHOMs generated, as well as their 
diversity? We implemented all introduced heuristics and calculated the rate of 
SSHOMs, as well as the diversity metric. We generate 1,000 HOMs for each 
heuristic, and repeat the process five times to remove the sampling bias.

\subsubsection*{RQ3. HOM Survival Rate} Which algorithm achieves the highest 
survival rate? We compared the survival rate of HOMs generated with each 
heuristic, along with the FOMs. The purpose of RQ3 is to see to what extent higher order 
mutants can survive compared to current First Order Mutation testing. For 
FOMs, we sample 1,000 mutants generated by MUSIC~\cite{Phan2018aa}.

\subsection{Diversity Metric}

We measure the diversity of mutant set by comparing the set of test 
cases that kill each mutant . For a mutant $m$ of original program $P$ with test 
suite $T$, a kill vector $v \in \mathcal{R}^{|T|}$ of $m$ is defined as 
follows.
\begin{equation*}
  v_i =
  \begin{cases}
    1
    &
    \text{if $T_i$ kills $m$;}
    \\[10pt]
    0
    &
    \text{otherwise.}
  \end{cases}
\end{equation*}
where $T_i$ refers to $i$-th test case of $T$. ($1 \le i \le |T|$)

The diversity of set of mutants $M = \{m_1, m_2, \dots, m_k\}$ is then defined as follows.

\begin{equation}
    dScore(M) = \frac{|\{v_1, v_2, \dots \}|}{|M|}
\end{equation}

\noindent where $v_i$ is the kill vector of $m_i$, and the numerator 
represents the number of distinct kill vectors. A higher $dScore$ is 
achieved if the mutant set contains more mutants with distinct kill 
vectors. The reason we define diversity with kill vectors is because two 
mutants with same kill vectors are not distinguishable from the point of test 
suite~\cite{Shin2016qy}.

\subsection{Benchmark Programs}

We study two C programs: a toy example called Bill's Car, and \texttt{schedule} from the SIR benchmark~\cite{Do:2005zp}.

\subsubsection{Bill's Car}

Algorithm~\ref{alg:billscar} shows Bill's Car, which is a C program that calculates parking fees. The fee depends on the 
day of the week, the car type, and minutes stayed in the parking zone.  There 
are three kinds of vehicle types. Senior, car, and truck. The fee structure, as 
well as discounts based on the day of week, means that the program contains nontrivial dependence structure. 


\begin{algorithm}[ht]
    \caption{Pseudo code of Bill's Car\label{alg:billscar}}
    \scriptsize
    \DontPrintSemicolon
    \SetKwProg{Main}{Function}{:}
    
    \Main{Main(vehicle, minutes, day)}{
        \eIf{$vehicle = senior$}
            {$fee = 0$}{
             \eIf{$vehicle \neq car$ \&\& $vehicle \neq truck$}
                {$Invalid Vehicle$}{
                \eIf{$vehicle = car$}{
                    $cost = Compute Car Fee(minutes)$}{
                    $cost = Compute Truck Fee(minutes)$
                }
                \eIf{$cost = -1$}
                    {$fee = -1$}{
                    \eIf{day = Thursday}
                        {$cost = 0.9 \times cost$}{
                        \If{day = Saturday}
                            {$cost = 1.1 \times cost$}
                    }
                        $fee = cost$ \\
                        $Print Fee(vehicle, day, minutes, fee)$
                }
            }
        }
    }
    
    \SetKwProg{Car}{Function}{:}
    
    \Car{ComputeCarFee(duration)}{
        $hours = duration / 60$ \\
        \eIf{$hours \le 2$}
            {$fee = 0$}{
            \eIf{$hours \le 5$}
                {$fee = 0.5 \times (hours - 2)$}{
                \eIf{$hours \le 15$}
                    {$fee = 0.5 \times 3 + 0.25 \times (hours - 5)$}{
                        $fee = -1$
                    }
            }
        }
    $return$ $fee$
    }
    
    \SetKwProg{Truck}{Function}{:}
    
    \Truck{ComputeTruckFee(duration)}{
        $hours = duration / 60$ \\
        \eIf{$hours \le 1$}
            {$fee = 0$}{
            \eIf{$hours \le 3$}
                {$fee = 1.0 \times (hours - 1)$}{
                \eIf{$hours \le 15$}
                    {$fee = 1.0 \times 2 + 0.75 \times (hours - 3)$}{
                        $fee = -1$
                    }
            }
        }
    $return$ $fee$    
    }
    
\end{algorithm}

The test suite for Bill's Car is consisted of 101 test cases and is constructed in 
a combinatorial manner. It covers three vehicle types, three categories of 
days of week, and 11 different time intervals (0 to 1,000 minutes at intervals 
of 100): this results in $3 \times 3 \times 11$ test cases. We add two edge 
cases, one with invalid car type and the other with missing arguments (only car type specified).
We use 6,400 mutants (100 mutants for each of 64 program elements) to build CPDA.

\subsubsection{Schedule}

Schedule is a C program from the Software-artifact Infrastructure 
Repository~\cite{Do:2005zp}. It is a schedule that calculates an ordering of 
given tasks. We use the coverage-extended test suite 456, which contains 81 
test cases. 
To build CPDA, we use 940 mutants (10 mutants for each of 94 program elements).

\subsection{Implementation and Environment}

Every first order mutants are made by C mutation testing tool, 
MUSIC~\cite{Phan2018aa}. Second order HOMs are generated by independently 
mutating program elements in the given pair, and combining the mutated lines.  
We use the \texttt{networkx} Python library to compute maximum weight matching.

Experiments for Bill's Car was ran on Ubuntu 18.04, Intel(R) Core(TM) i7-10700 
CPU @ 2.90GHz with GeForce RTX 3070. Experiments for schedule was ran on 
Ubuntu 16.04.5, Intel(R) Core(TM) i7-6700 CPU @ 3.40GHz with GeForce GTX 1080.

%% file: Result.tex
\section{Results}
\label{sec:results}


\begin{table}[t]
\caption{Causal Effect and Average Number of SSHOMs per Bucket}\label{tbl:buckets}
\centering
\scalebox{0.8}{
\begin{tabular}{r||rrr|rrr}
 \toprule
 & \multicolumn{3}{c|}{Bill's Car} & \multicolumn{3}{c}{Schedule} \\ 
 Buc. & Pairs & CE Range & Avg. SSHOMs & Pairs & CE Range  & Avg. SSHOMs\\ \midrule
 0  & 2,094 & 0             & 0.1  & 1,392 & 0             & 0.6 \\
 1  &       & 0.004 - 0.144 & 0.0  &       & 0.008 - 0.021 & 0.0 \\
 2  &       & 0.145 - 0.229 & 0.0  &       & 0.021 - 0.030 & 0.1 \\
 3  &       & 0.229 - 0.297 & 0.0  &       & 0.030 - 0.051 & 1.8 \\
 4  &       & 0.297 - 0.349 & 0.0  &       & 0.051 - 0.080 & 3.7 \\
 5  &       & 0.349 - 0.397 & 0.0  &       & 0.080 - 0.133 & 0.1 \\
 6  &       & 0.397 - 0.437 & 1.1  &       & 0.133 - 0.167 & 5.4 \\
 7  &       & 0.437 - 0.486 & 1.7  &       & 0.167 - 0.218 & 0.0 \\
 8  &       & 0.487 - 0.553 & 0.7  &       & 0.222 - 0.322 & 1.9 \\
 9  &       & 0.553 - 0.669 & 1.1  &       & 0.326 - 0.495 & 4.1 \\
 10 &       & 0.669 - 1.000 & 12.9 &       & 0.495 - 0.997 & 11.8\\
 \bottomrule
\end{tabular}}
\end{table}


\subsection{RQ1: Causal Effect and number of SSHOM}

Table~\ref{tbl:buckets} shows the results of the bucketing analysis for RQ1. 
While Bill's Car had more program element pairs with zero causal effect, 
Schedule had more program element pairs with positive causal effects.  
For Bill's Car, $2.25\%$ of the total pair of program elements turned out to 
have causal effect value over $0.5$ while for Schedule $18\%$ of pairs were. 
The range of causal effect values for each bucket tends to increase. 
Average range of first three buckets for the two benchmark programs was 
$0.056$, while the average range of final three buckets was $0.214$. 
This suggests that the space of program pairs get sparse as the causal effect 
value goes up.

\begin{figure}[ht]
\centering
\subfigure[Bill's car\label{fig:SSR-bill}]{\includegraphics[angle=0, width=0.23\textwidth]{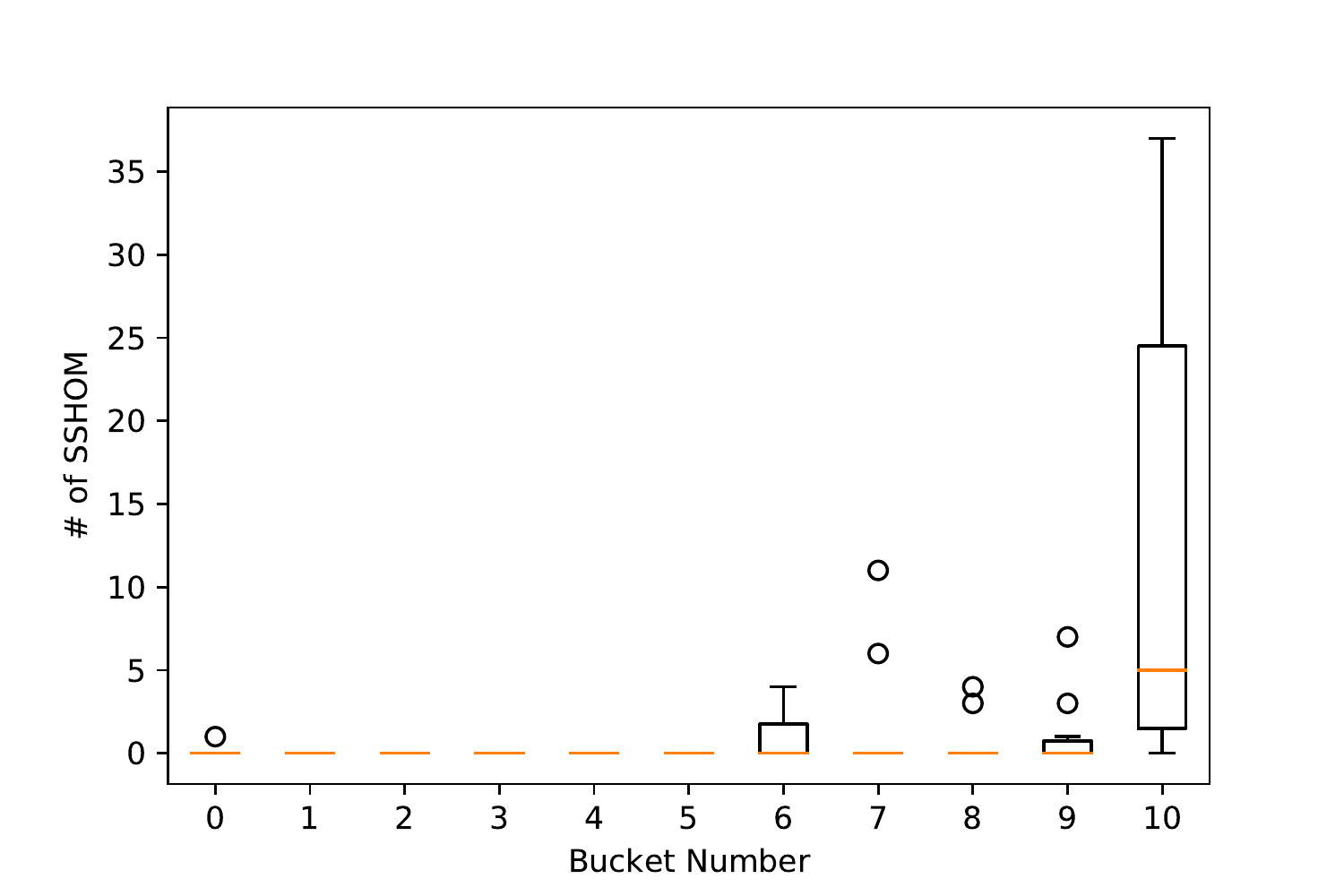}}
\subfigure[Schedule\label{fig:SSR-sch}]{\includegraphics[angle=0, width=0.23\textwidth]{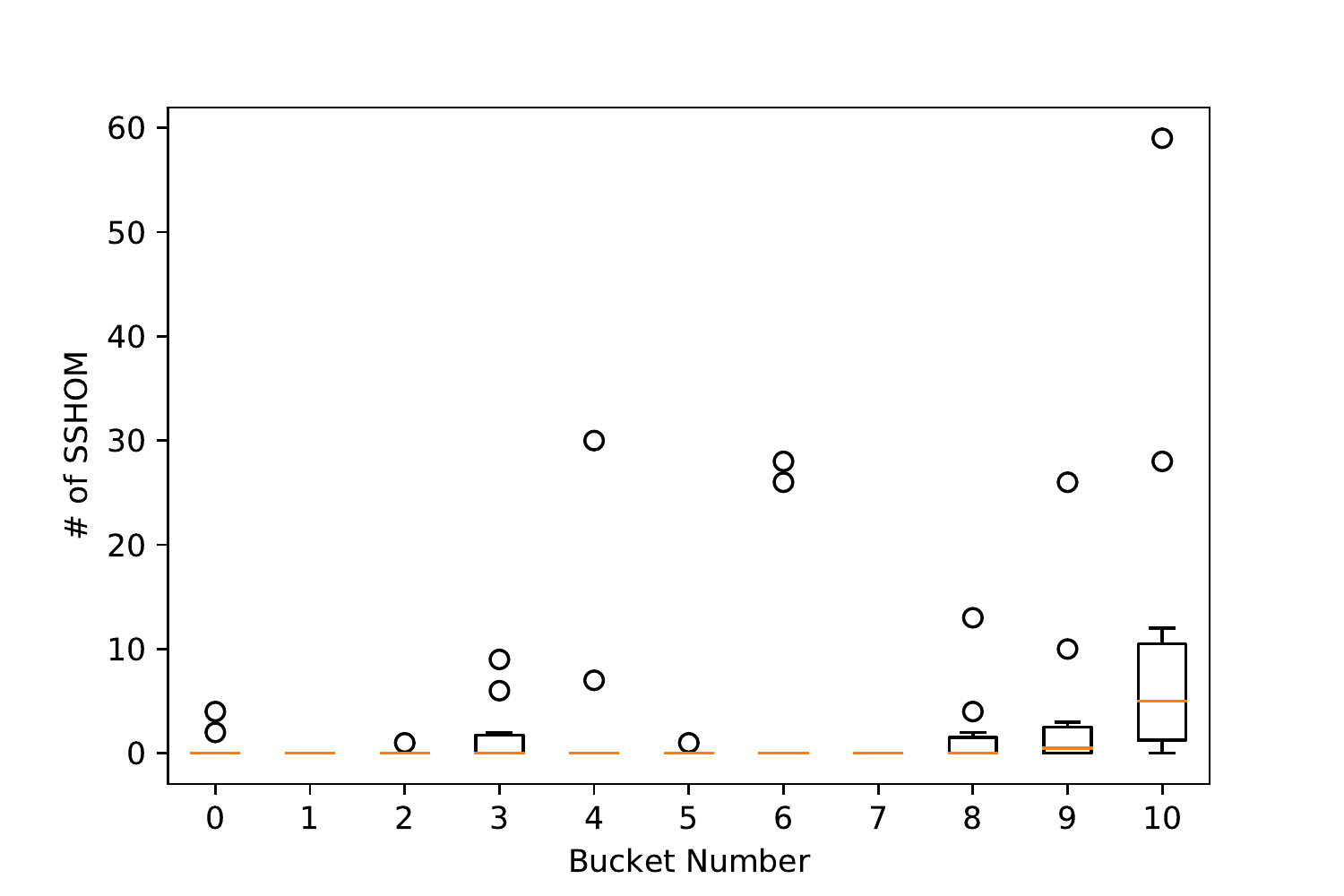}}
\caption{Boxplot of Number of SSHOMs per Bucket\label{fig:boxplots}}
\end{figure}

Figure~\ref{fig:SSR-bill} and \ref{fig:SSR-sch} shows the number of SSHOM for 
the two benchmark programs. From all SSHOM made from all trials, $73.71\%$ 
SSHOM were from the top bucket for Bill's Car while for Schedule, $40.83\%$ 
SSHOM were. Specific number of average SSHOM made per bucket for each trial (5 
pairs from bucket with 100 HOMs each) is shown in Table~\ref{tbl:buckets}. We 
were able to observe that the top bucket does significantly better job of 
generating SSHOM than other buckets. 

Since number of program element pairs in the top bucket are much smaller than 
total pairs ($6.85$\% for Schedule, $2.40$\% for Bill's Car), we can 
significantly reduce the search space for second order HOMs by focusing on 
the pairs in the top bucket. Based on these results, we conclude that mutating 
program elements with high causal effect can lead to the generation of 
SSHOMs with higher probability.

\begin{figure}[ht]
\centering
\subfigure[Bill's Car - Number of SSHOMs\label{fig:billssr}]{\includegraphics[angle=0, width=0.24\textwidth]{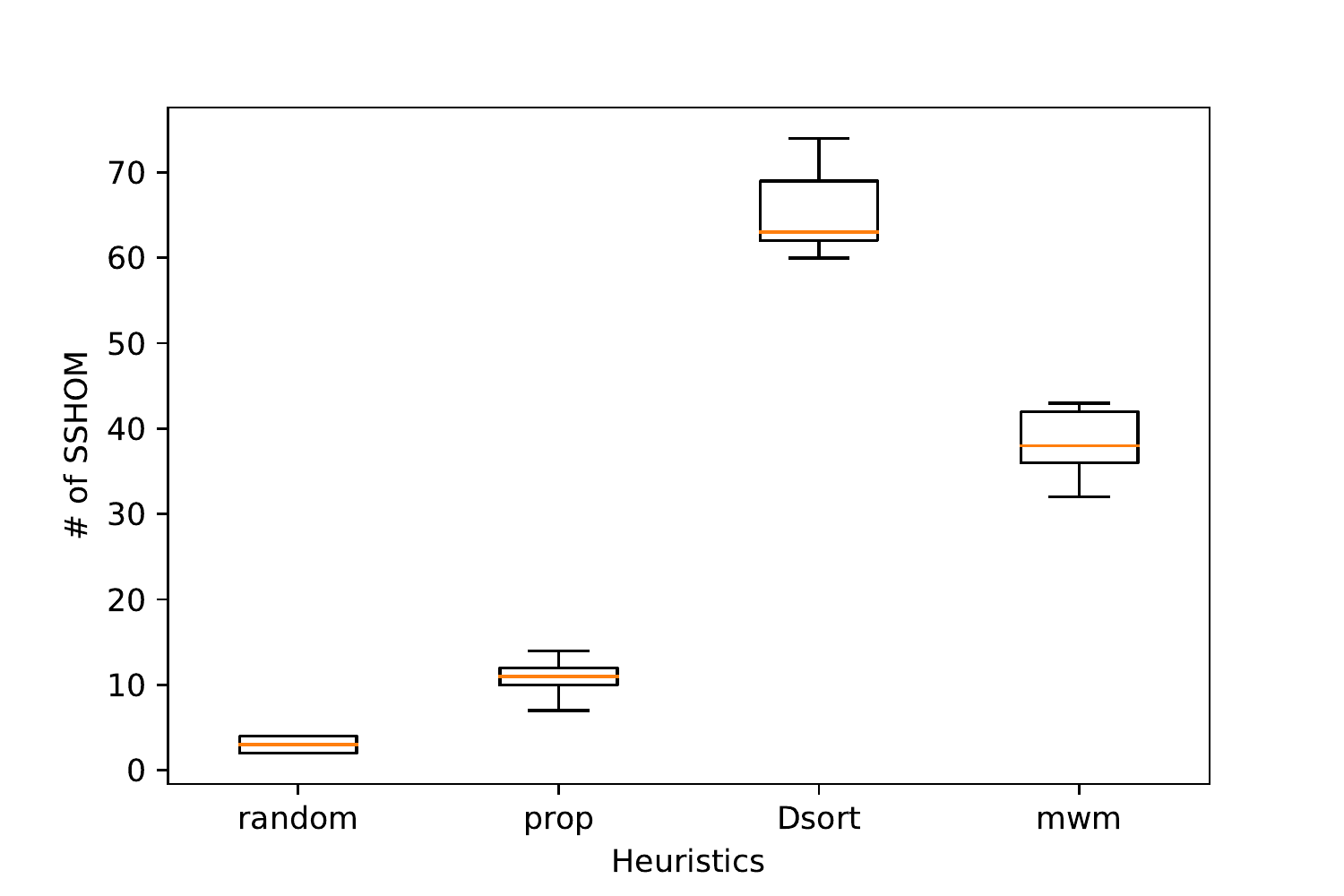}}
\subfigure[Bill's Car - Unique SSHOMs\label{fig:billssmdiv}]{\includegraphics[angle=0, width=0.24\textwidth]{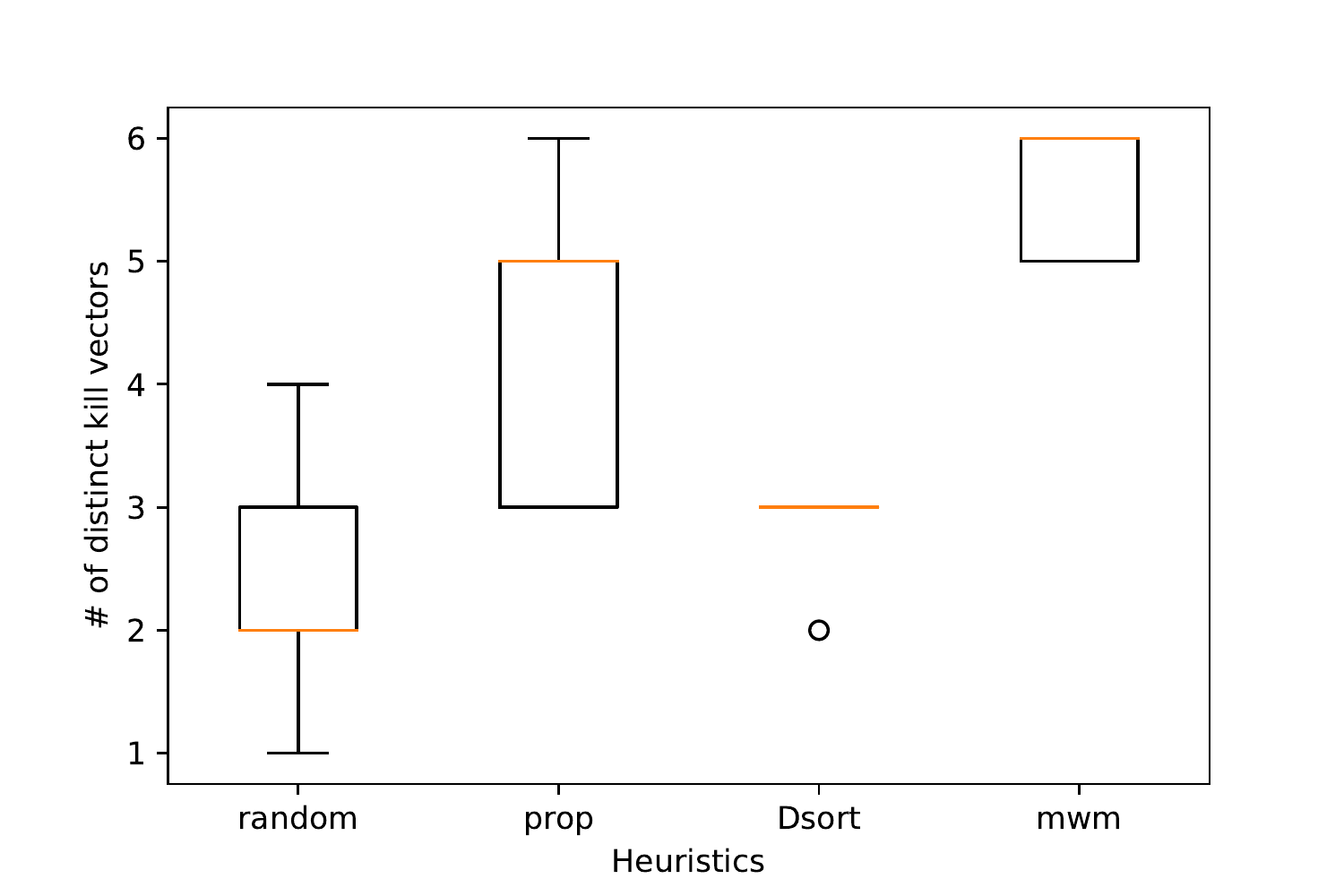}}
\subfigure[Schedule - Number of SSHOMs\label{fig:schedulessr}]{\includegraphics[angle=0, width=0.24\textwidth]{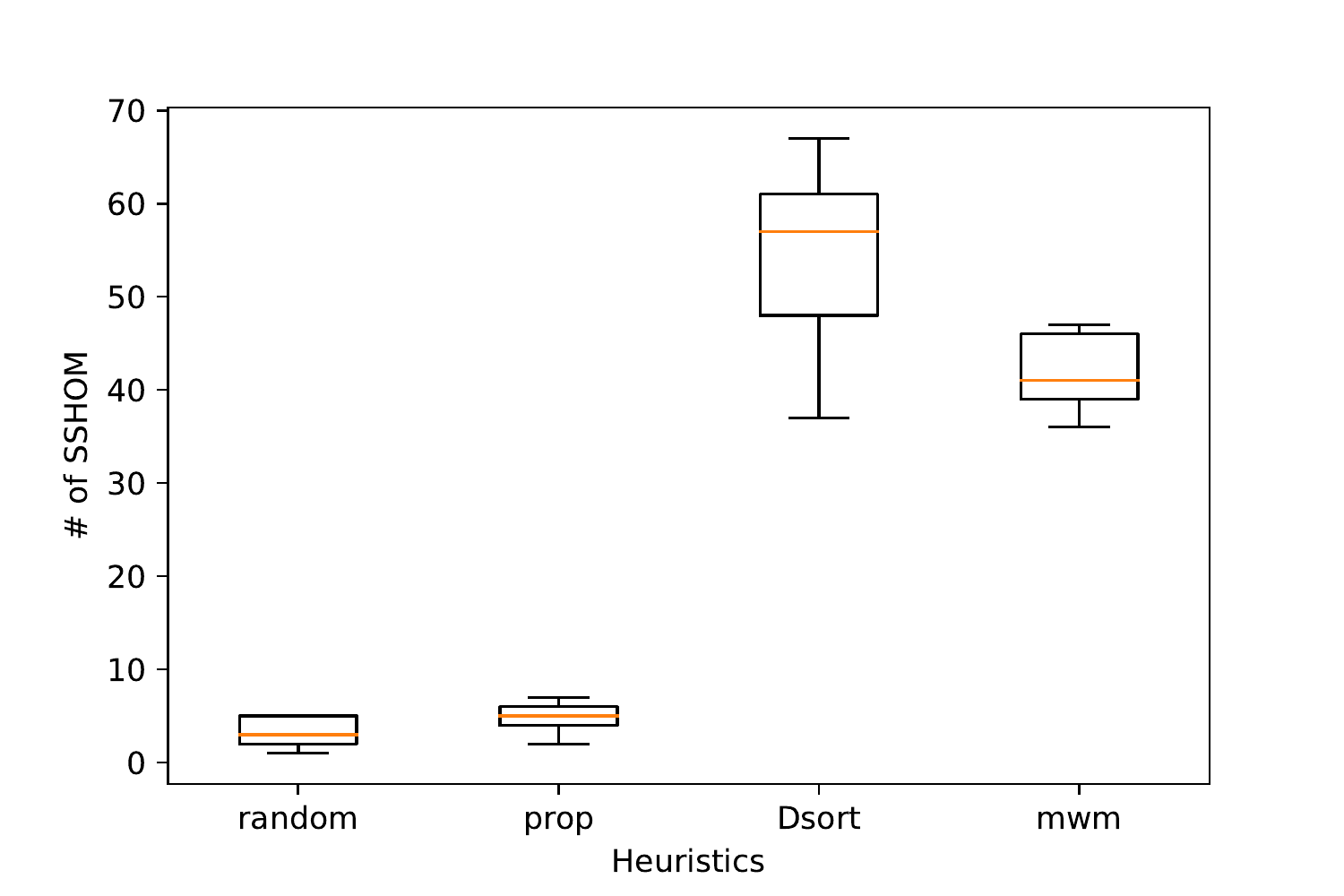}}
\subfigure[Schedule - Unique SSHOMs\label{fig:schedulessmdiv}]{\includegraphics[angle=0, width=0.24\textwidth]{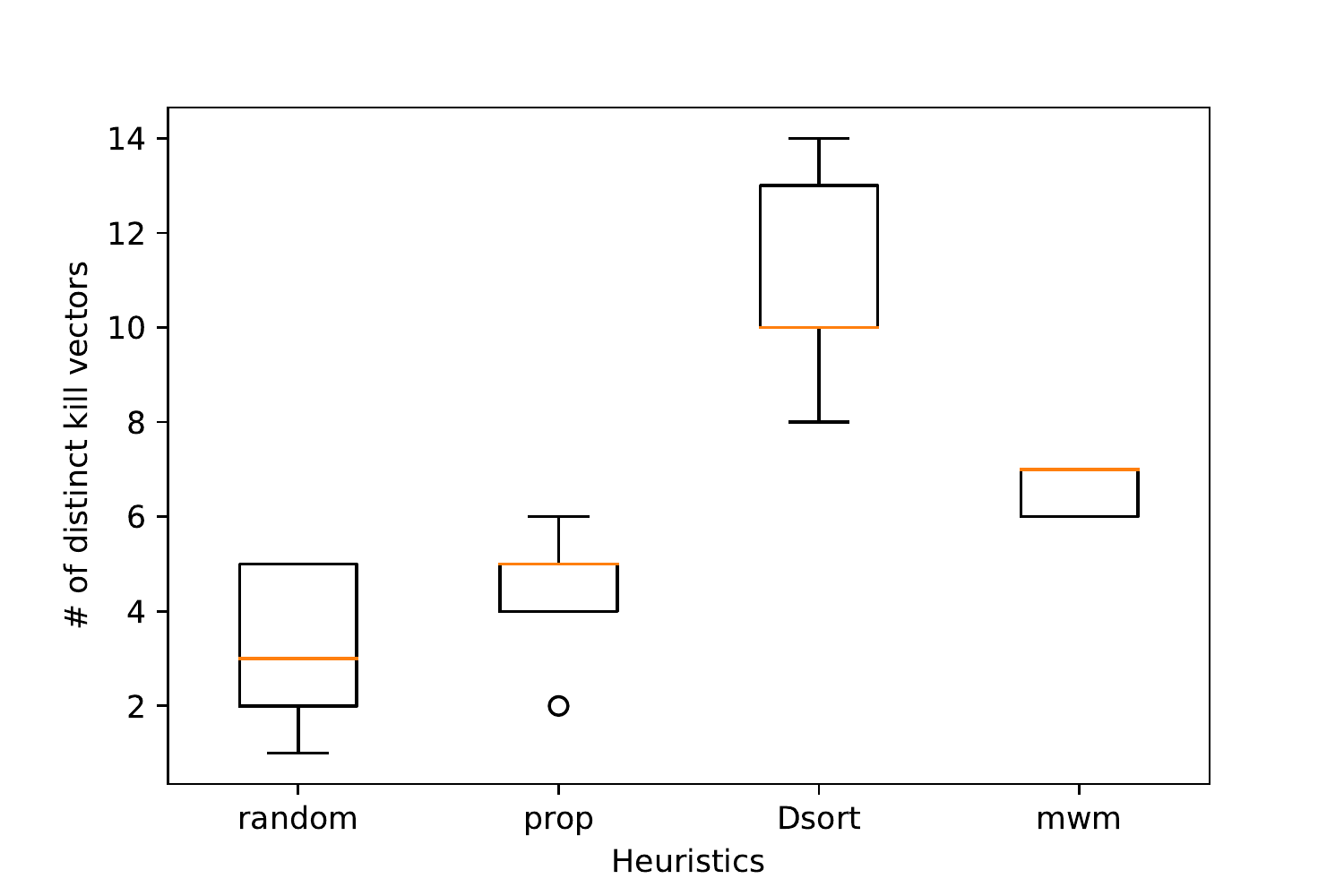}}
\caption{Number of Total and Unique SSHOMs Generated by Heuristics}
\label{fig:rq2total}
\end{figure}

\begin{table}[ht]
\centering
\caption{Mutant Diversity of Heuristics}
\label{tbl:rq2}
\scalebox{0.8}{
\begin{tabular}{lrrr|lrrr}
\toprule
Heuristic & \multicolumn{3}{c}{Bill's Car} & \multicolumn{3}{c}{Schedule}  \\
          & dScore                         & SSHOM                            & Uniq. SSHOM & dScore & SSHOM & Uniq. SSHOM \\ \midrule
Random    & 0.282                          & 3.0                              & 2.4         & 0.457  & 3.2   & 3.2         \\
Prop      & 0.226                          & 10.8                             & 4.4         & 0.471  & 4.8   & 4.4         \\
Dsort     & 0.096                          & 65.6                             & 2.8         & 0.135  & 54.0  & 11.0        \\
MWM       & 0.118                          & 38.2                             & 5.6         & 0.282  & 41.8  & 6.6         \\
\bottomrule
\end{tabular}}
\end{table}


\subsection{RQ2: Performance of each Heuristic}
Figure~\ref{fig:rq2total} shows for each benchmark programs the number of 
evaluated SSHOM, dScore, and the number of distinct kill vectors for the 
generated SSHOM. Table~\ref{tbl:rq2} shows the average number of calculated 
metrics. Column dScore contains the diversity score of all generated HOMs (not 
necessarily strongly subsuming); column SSHOM contains the average number of 
generated SSHOMs, and column 
Uniq. SSHOM contains the average number of generated SSHOMs with unique kill 
vectors. Dsort generates the largest number of SSHOMs, which is $21.87$, 
and $16.88$ times more than Random for Bill's Car and Schedule, respectively. 
Although MWM produced fewer SSHOMs, it still generates $12.73$, and $13.06$ 
times more SSHOM than Random, respectively. This reflects the fact that Dsort 
only prioritises the selection of pairs based causal effects. 

Prop is less successful in generating many SSHOMs: by definition 
it ends up choosing more program pairs than Dsort or MWM, resulting in 
generation of fewer mutants per chosen pair. For Schedule, Prop selects 818.8 
pairs on average, while Random selects 897.8, and Dsort and MWM only 21. 
However, the higher diversity of chosen pairs results in higher dScore.

\begin{figure}[t]
\centering
\subfigure[MWM\label{fig:chosen_mwm}]{\includegraphics[angle=0, width=4cm]{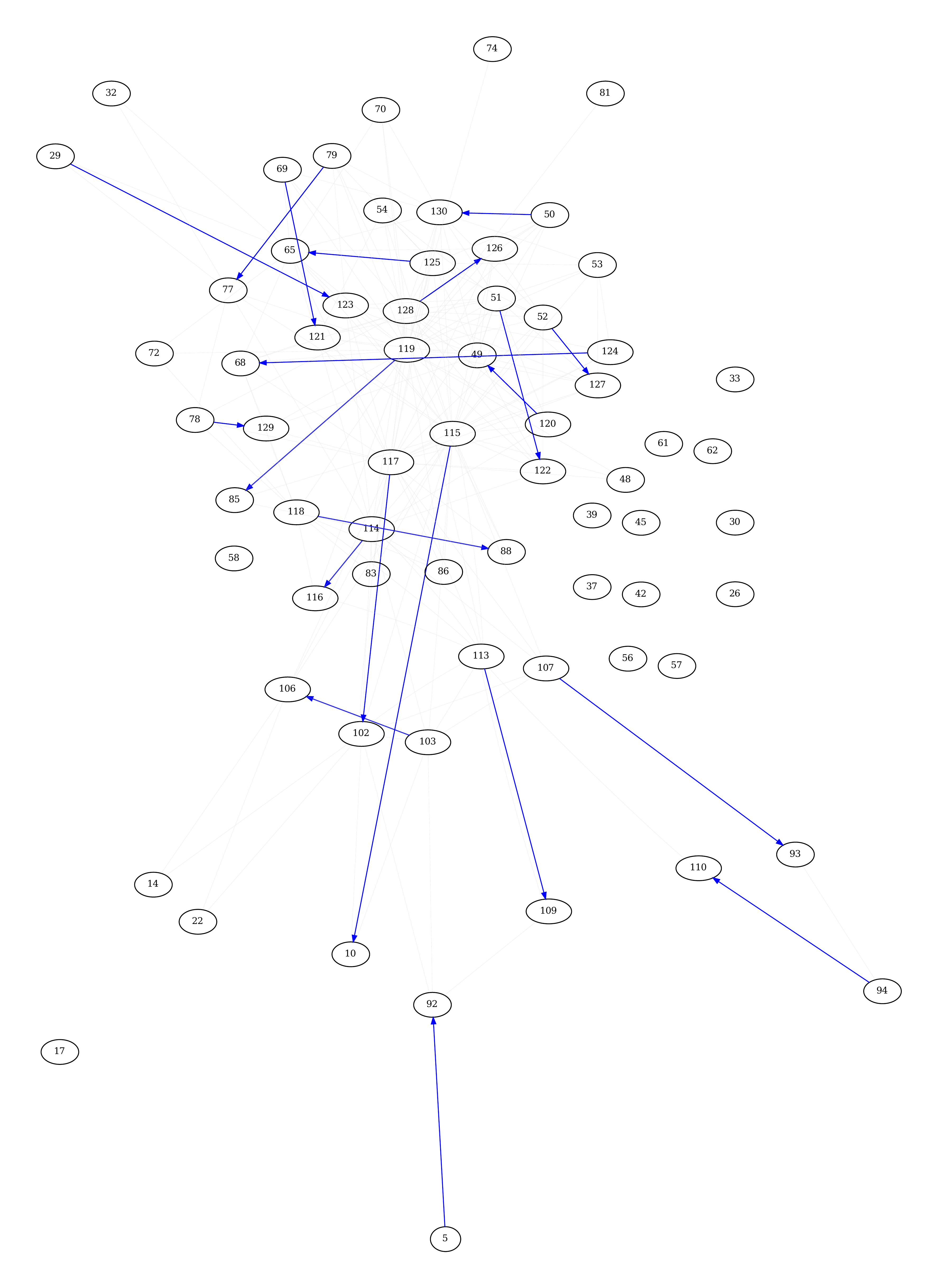}}
\subfigure[Dsort\label{fig:chosen_dsort}]{\includegraphics[angle=0, width=4cm]{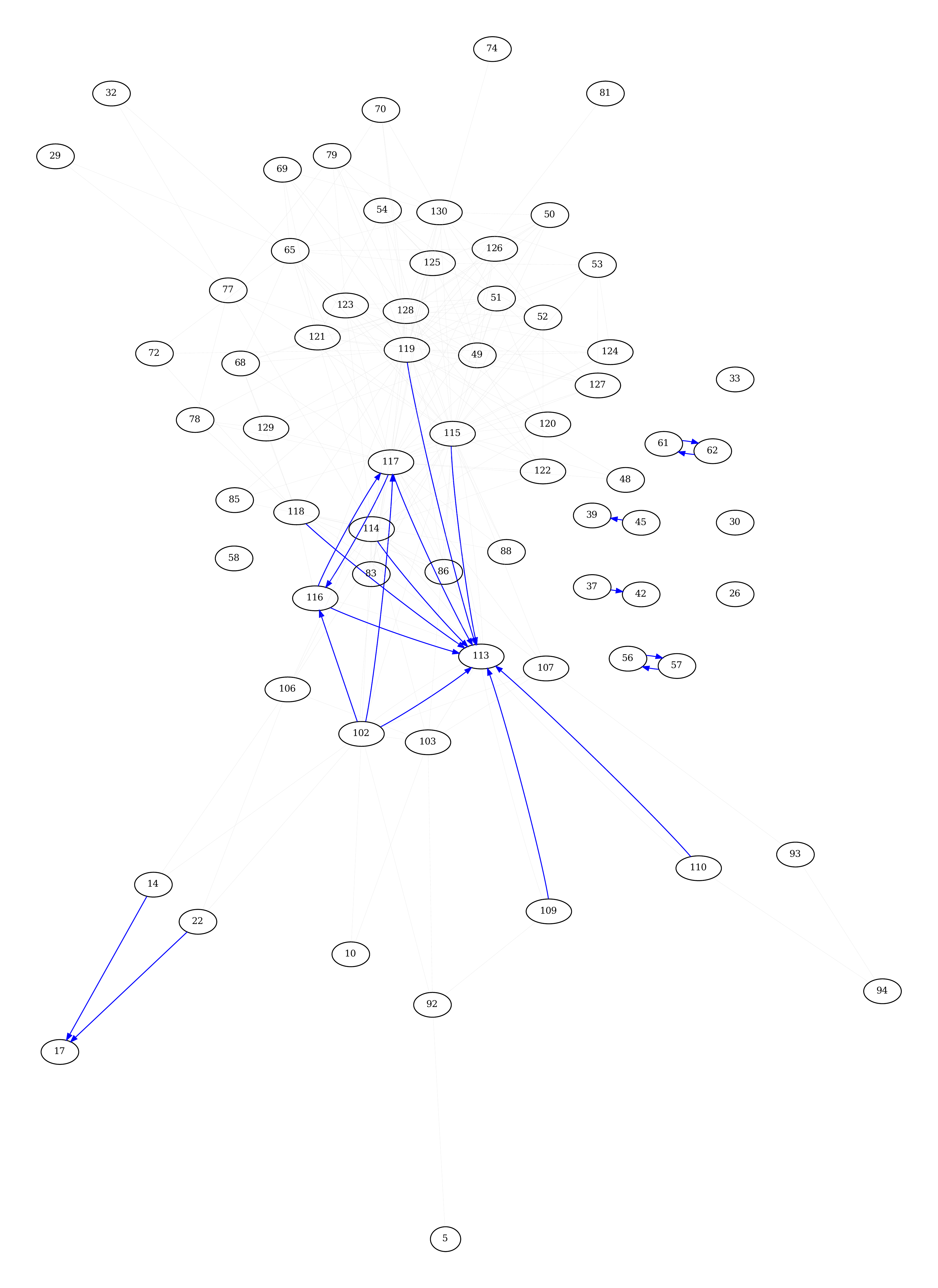}}
\caption{Pairs of Program Elements in Schedule Chosen by MWM and Dsort\label{fig:chosen_pairs}}
\end{figure}



Figure~\ref{fig:billssmdiv} and \ref{fig:schedulessmdiv} shows boxplots of the 
unique number of SSHOMs generated by different heuristics. MWM produces more unique SSHOMs for Bill's Car than Dsort, but the 
trend is the opposite in Schedule. We suspect that diversity of HOMs in 
general (captured by dScore in Table~\ref{tbl:rq2}), and the diversity of 
SSHOMs, may not align perfectly. 

\begin{figure}[ht]
\centering
\subfigure[Bill's Car - \# of Survived Mutants\label{fig: Bill's Car Comparison with FOM}]{\includegraphics[angle=0, width=0.24\textwidth]{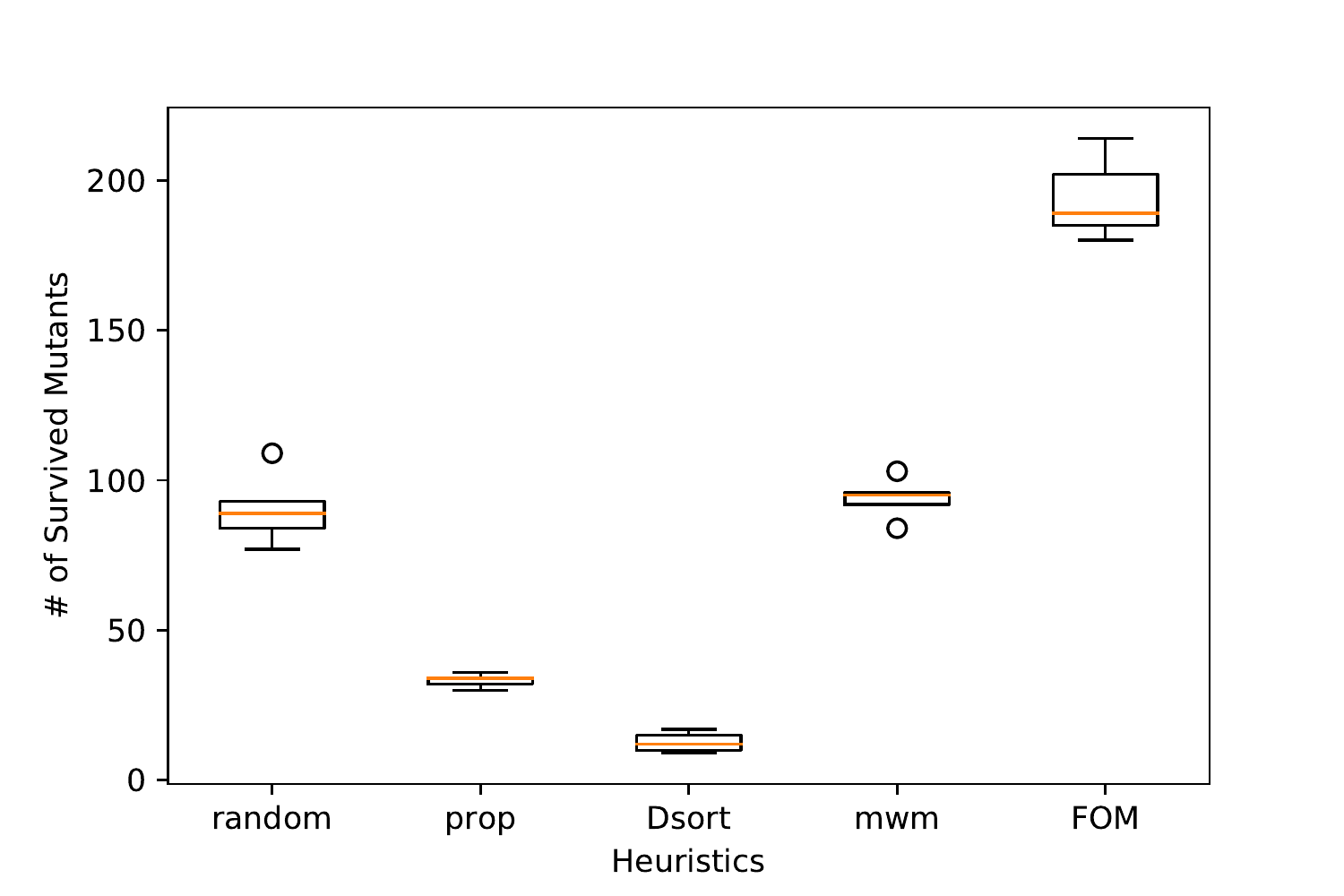}}
\subfigure[Schedule - \# of Survived Mutants\label{fig: Schedule Comparison with FOM}]{\includegraphics[angle=0, width=0.24\textwidth]{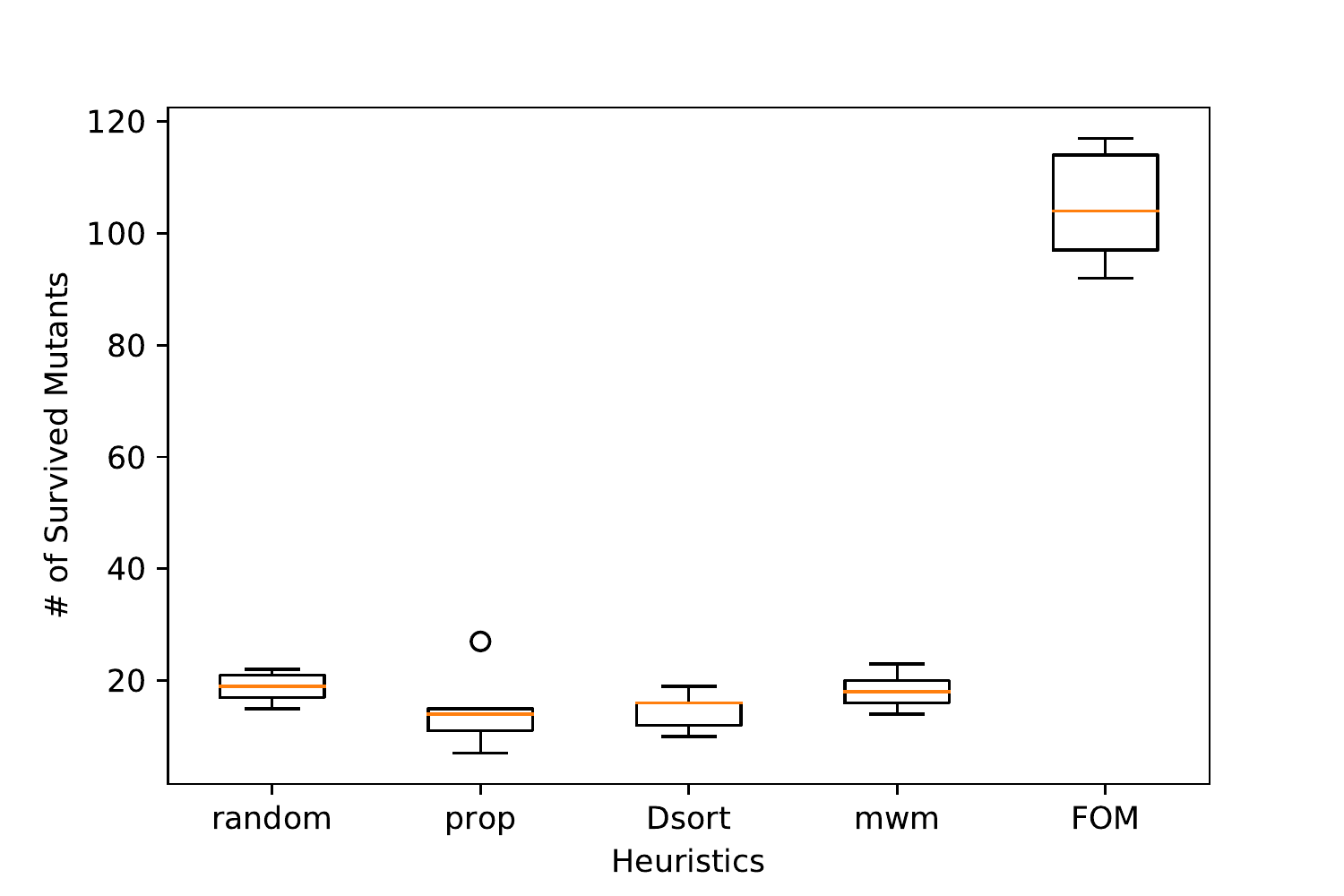}}
\caption{Survived Mutants}\label{fig:survived}    
\end{figure}

MWM heuristic successfully diversified the generated mutants. Figure~\ref{fig:chosen_mwm} and \ref{fig:chosen_dsort} visualises the program element 
pairs chosen by MWM and Dsort. The causal structure is shown in gray edges, 
while chosen pairs are shown with blue arrowed edges. While chosen pairs from 
MWM tend to be spread out in various places by not sharing common vertices, 
pairs from Dsort does overlap and tend to cover only some specific regions of 
the program, as expected.

\begin{table}[ht]
\centering
\caption{Average number of Survived Mutants\label{tbl:avgsurvive}}
\scalebox{0.8}{
\begin{tabular}{lrr}
\toprule
       & Bill's Car & Schedule \\ \midrule
Random & 90.4       & 18.8     \\ 
Prop   & 33.2       & 14.8     \\ 
Dsort  & 12.6       & 14.6     \\ 
MWM    & 94.0       & 18.2     \\ 
FOM    & 194.0      & 104.8    \\ 
\bottomrule
\end{tabular}}
\end{table}

\subsection{RQ3: Survival Rate of each Heuristic}

Figure~\ref{fig:survived} shows the number of surviving mutants generated by 
each heuristic, along with the number of surviving FOMs. The average values 
are reported in Table~\ref{tbl:avgsurvive}. It shows that HOMs are easier to kill, possibly due to the larger semantic differences. 

Among the proposed heuristics, MWM showed highest number of surviving mutants, 
followed by Random. Dsort shows the lowest survival rate. The survival rate 
differs a lot for Bill's Car while there are not so much variance in Schedule.

We observe that, in Bill's Car, there are specific program locations that 
produce more surviving mutants than others. For example, mutants generated in 
the flow of \texttt{PrintFee} function tend to survive more. Since
the main objective of the function is to print status, the return value is not 
used anywhere. Consequently, it is harder to kill.

We also observe that pairs with high causal effect values tend to exist in the 
part of program with main functionalities. For example, pairs chosen by Dsort  
from Bill's Car are mostly from functions calculating the fee, while MWM also 
chooses from the \texttt{PrintFee} function, resulting in a higher survival 
rate. Random and Prop all showed high survival rate due to a similar reason. 
In Schedule, survival rates of mutants are not significantly affected by the 
location of mutation.

%% file: related_work.tex

\section{Related Work}
\label{sec:related_work}

The concept of Subsuming Higher Order Mutants was proposed by Jia et 
al.~\cite{Jia2009id}, as a way to avoid equivalent mutants and to reduce the 
number of mutants to examine. Jia et al. present more detailed classification 
of HOMs, but we focus only on SSHOMs in this preliminary study.

One of the most widely studied topic in Higher Order Mutation Testing is how 
to efficiently generate SSHOMs. Harman et al. generates SSHOMs using genetic
algorithm~\cite{Harman14ase}. Since the fitness evaluation involves executing 
all candidate SSHOMs, the cost of the search-based approach can be high. Wong 
et al.~\cite{Wong2020dm} uses variational execution and SAT solver to 
efficiently find SSHOMs. Our approach depends on CPDA, which in turn uses 
mutation analysis to compute concrete causal effect values~\cite{Lee:2021aa}. 
However, compared to fitness guided search, the cost of CPDA can be controlled 
by the parameters (i.e., how many mutants to consider for CPDA). 


%% file: Conclusion.tex

\section{Conclusion}
\label{sec:conclusion}

We propose a new approach of sampling higher order mutants by using Causal 
Program Dependence Analysis (CPDA). Specifically, we show that causal effect 
can effectively guide the generation of SSHOMs. We compare four different 
SSHOM generation heuristics and the SSHOMs generated by them. The quality of 
mutant set is measured in terms of two metrics, number of SSHOM and 
diversity. Our results show that MWM and Dsort heuristics 
can effectively sample SSHOMs. For future work, we plan to add more benchmark 
programs, and investigate more sophisticated heuristic design that considers factors other than causal effects simultaneously.

%% file: main.bbl
\begin{thebibliography}{10}
\providecommand{\url}[1]{#1}
\csname url@samestyle\endcsname
\providecommand{\newblock}{\relax}
\providecommand{\bibinfo}[2]{#2}
\providecommand{\BIBentrySTDinterwordspacing}{\spaceskip=0pt\relax}
\providecommand{\BIBentryALTinterwordstretchfactor}{4}
\providecommand{\BIBentryALTinterwordspacing}{\spaceskip=\fontdimen2\font plus
\BIBentryALTinterwordstretchfactor\fontdimen3\font minus
  \fontdimen4\font\relax}
\providecommand{\BIBforeignlanguage}[2]{{%
\expandafter\ifx\csname l@#1\endcsname\relax
\typeout{** WARNING: IEEEtran.bst: No hyphenation pattern has been}%
\typeout{** loaded for the language `#1'. Using the pattern for}%
\typeout{** the default language instead.}%
\else
\language=\csname l@#1\endcsname
\fi
#2}}
\providecommand{\BIBdecl}{\relax}
\BIBdecl

\bibitem{Jia:2011nx}
Y.~Jia and M.~Harman, ``An analysis and survey of the development of mutation
  testing,'' \emph{IEEE transactions on software engineering}, vol.~37, no.~5,
  pp. 649--678, 2011.

\bibitem{Offutt:1992}
\BIBentryALTinterwordspacing
A.~J. Offutt, ``Investigations of the software testing coupling effect,''
  \emph{ACM Trans. Softw. Eng. Methodol.}, vol.~1, no.~1, pp. 5--20, Jan. 1992.
  [Online]. Available: \url{https://doi.org/10.1145/125489.125473}
\BIBentrySTDinterwordspacing

\bibitem{Madeyski2014}
L.~{Madeyski}, W.~{Orzeszyna}, R.~{Torkar}, and M.~{J{\'o}zala}, ``Overcoming
  the equivalent mutant problem: A systematic literature review and a
  comparative experiment of second order mutation,'' \emph{IEEE Transactions on
  Software Engineering}, vol.~40, no.~1, pp. 23--42, 2014.

\bibitem{Jia2009id}
Y.~Jia and M.~Harman, ``Higher order mutation testing,'' \emph{Information and
  Software Technology}, vol.~51, no.~10, pp. 1379--1393, 2009.

\bibitem{Harman14ase}
\BIBentryALTinterwordspacing
M.~Harman, Y.~Jia, P.~Reales~Mateo, and M.~Polo, ``Angels and monsters: An
  empirical investigation of potential test effectiveness and efficiency
  improvement from strongly subsuming higher order mutation,'' in
  \emph{Proceedings of the 29th ACM/IEEE International Conference on Automated
  Software Engineering}, ser. ASE '14.\hskip 1em plus 0.5em minus 0.4em\relax
  New York, NY, USA: Association for Computing Machinery, 2014, p. 397–408.
  [Online]. Available: \url{https://doi.org/10.1145/2642937.2643008}
\BIBentrySTDinterwordspacing

\bibitem{Lee:2021aa}
S.~Lee, D.~Binkley, R.~Feldt, N.~Gold, and S.~Yoo, ``Causal program dependence
  analysis and causal fault localization,'' Korea Advanced Institute of Science
  and Technology, 291 Daehak-ro, Yuseong-gu, Daejeon, Korea 34141, Tech. Rep.
  CS-TR-2021-423, January 2021.

\bibitem{Pearl:2009aa}
J.~Pearl, \emph{Causality}.\hskip 1em plus 0.5em minus 0.4em\relax Cambridge
  University Press, 2009.

\bibitem{Diestel:2016aa}
R.~Diestel, \emph{Graph Theory}, 5th~ed., Springer-Verlag, Heidelberg, August
  2016, vol. 173.

\bibitem{Phan2018aa}
D.~L. {Phan}, Y.~{Kim}, and M.~{Kim}, ``Music: Mutation analysis tool with high
  configurability and extensibility,'' in \emph{Proceedings of IEEE
  International Conference on Software Testing, Verification and Validation
  Workshops (ICSTW)}, ser. Mutation 2018, April 2018, pp. 40--46.

\bibitem{Shin2016qy}
D.~Shin and D.~H. Bae, ``A theoretical framework for understanding
  mutation-based testing methods,'' in \emph{2016 IEEE International Conference
  on Software Testing, Verification and Validation (ICST)}, April 2016, pp.
  299--308.

\bibitem{Do:2005zp}
H.~Do, S.~G. Elbaum, and G.~Rothermel, ``Supporting controlled experimentation
  with testing techniques: An infrastructure and its potential impact.''
  \emph{Empirical Software Engineering}, vol.~10, no.~4, pp. 405--435, 2005.

\bibitem{Wong2020dm}
\BIBentryALTinterwordspacing
C.-P. Wong, J.~Meinicke, L.~Chen, J.~a.~P. Diniz, C.~K\"{a}stner, and
  E.~Figueiredo, ``Efficiently finding higher-order mutants,'' in
  \emph{Proceedings of the 28th ACM Joint Meeting on European Software
  Engineering Conference and Symposium on the Foundations of Software
  Engineering}.\hskip 1em plus 0.5em minus 0.4em\relax New York, NY, USA:
  Association for Computing Machinery, 2020, pp. 1165--1177. [Online].
  Available: \url{https://doi.org/10.1145/3368089.3409713}
\BIBentrySTDinterwordspacing

\end{thebibliography}
